\documentclass[11pt,a4paper]{article}

\usepackage{amsfonts}
\usepackage{amsmath}
\usepackage{amssymb}
\usepackage{amsthm}
\usepackage{lscape}

\usepackage[active]{srcltx}
\usepackage{ifpdf}
\usepackage{setspace}

\usepackage[font=footnotesize,justification=raggedright,parskip=5pt]{caption,
subfig}

\ifpdf

\newcommand{\drivy}{pdftex}
\usepackage[pdftex=true,hyperfootnotes=true]{hyperref}
\else

\newcommand{\drivy}{dvips}
\usepackage[hyperfootnotes=true]{hyperref}
\fi

\hypersetup{colorlinks=true,raiselinks=true,breaklinks=false}

\usepackage[\drivy]{graphicx}
\usepackage{natbib}
\newtheorem{thm}{Theorem}

\theoremstyle{definition}

\theoremstyle{remark}

\newcommand{\abs}[1]{\left\vert#1\right\vert}

\newcommand{\D}{\mathfrak{D}}

\newcommand{\T}{\top}

\newcommand{\brac}[1]{\left ( #1 \right )}
\newcommand{\brak}[1]{\left [ #1 \right ]}
\newcommand{\brat}[1]{\left \{ #1 \right \}}
\newcommand{\Qmeas}{\mathbb{Q}}
\newcommand{\QQ}{\mathbb{Q}}
\newcommand{\Pmeas}{\mathbb{P}}
\newcommand{\PP}{\mathbb{P}}

\newcommand{\evt}[2]{\mathbb{E}^{\Pmeas}_{#1}\brak{#2}}
\newcommand{\evqt}[2]{\mathbb{E}^{\Qmeas}_{#1}\brak{#2}}
\newcommand{\evind}[2]{\mathbb{E}_{#1}\brak{#2}}

\newcommand{\kq}{b_1 ^{\Qmeas}}

\newcommand{\kthq}{b_0 ^{\Qmeas}}

\newcommand{\model}{\mathcal{M}}

\newcommand{\reals}{\mathbb{R}}
\newcommand{\RR}{\mathbb{R}}

\newcommand{\dblaw}[2]{\mathbb{Q}_\model ^{#1, #2}}
\newcommand{\dbrlaw}[2]{\mathbb{W}_\model ^{#1, #2}}
\newcommand{\lsum}[2]{ \sum\limits _{#1}^{#2}}
\usepackage[latin1]{inputenc}
\title{Empirical Asset Pricing with Nonlinear Risk Premia\thanks{We are grateful to Yacine A\"{i}t-Sahalia, Valentina Corradi, and Damir Filipovi\'{c} for insightful
comments.}}
\author{Aleksandar Mijatovi\'{c}\thanks{Department of Mathematics,
Imperial College London, Huxley Building, 180 Queen's Gate, London SW7 2AZ, United Kingdom,
\href{mailto:a.mijatovic@imperial.ac.uk}{\texttt{a.mijatovic@imperial.ac.uk}}}
\and Paul Schneider\thanks{Finance Group, 
Warwick Business School, University of Warwick, Coventry CV4 7AL, United Kingdom,
\href{mailto:paul.schneider@wbs.ac.uk}{\texttt{paul.schneider@wbs.ac.uk}}}}

\begin{document}
\maketitle
\begin{abstract}
In this paper we
introduce a simple continuous-time asset pricing framework, based on general
multi-dimensional
diffusion processes, that combines semi-analytic pricing with 
a nonlinear specification for the market price of risk.
Our framework guarantees existence of weak solutions of the nonlinear
SDEs under the physical measure, thus allowing
to work with nonlinear models for the real world dynamics 
not considered in the literature so far. 
It emerges that
the additional flexibility 
in the time series modelling is econometrically 
relevant: a nonlinear stochastic volatility diffusion model for the joint time series of the S\&P 100 and the VXO implied volatility 
index data shows superior forecasting power over the standard specifications for implied and realized variance forecasting.
\end{abstract}




\section{Introduction}\label{sec:intro}
Most financial time series exhibit rapid fluctuations while being extremely persistent at the same time. 
Violent fluctuations are often identified as jumps caused by events such as central bank meetings or rating announcements. 
The economic intuition suggests that for example interest rates should be stationary. However
unit-root tests often imply that interest rates are integrated and therefore exhibit extreme persistence.
Ideally a model should be able to accommodate both extremes while maintaining 
compatibility with economic theory:
random walk like behaviour in a certain region, and reversion towards a mean  
outside it.
At first glance establishing the existence of such a model under the
real world measure appears to be very difficult. 
A diffusion process with these characteristics would clearly 
need to exhibit a highly nonlinear drift under the physical measure,
which implies that global Lipschitz and growth conditions, 
typically required
for the existence of a solution to a multi-dimensional SDE, 
are not satisfied. 
In a univariate diffusion setting~\citet{aitsahalia96} 
applies a more general method, only available in dimension one, to ascertain the existence of a model
that exhibits the desired characteristics.
The reason for the econometric success of his model lies in the  nonlinearity 
of the drift.
Two main obstacles to a wide applicability of such models
remain.
The first is the lack of closed-form, or at least semi-analytic, solutions for the prices 
of contingent claims within the nonlinear framework. 
The second is a lack of tools for proving the existence of solutions 
to the stochastic differential equations used when attempting to introduce 
nonlinearity in a multivariate setting.

Employing econometrically inconspicuous dampening functions we introduce a 
simple multivariate diffusion framework which
exploits the existence of a solution of an SDE
under a risk-neutral probability measure and guarantees the existence
of a weak solution of a nonlinear SDE under the real world probability 
measure.
From the econometric point of view
our framework  extends the affine approach 
from~\citet{cheriditofilipovickimmel03} 
yielding substantially enriched dynamics.
The most obvious application is a state variable formulation that entails 
(semi-)analytic pricing under the risk-neutral measure, 
which leaves flexibility for the dynamics 
under the physical measure similar to that of the discrete-time approach  
considered in~\citet{dailesingleton06} and~\citet{bertholonmonfortpegoraro08}.  
Recent advances in estimating the parameters of nonlinear diffusions such as the 
algorithms introduced in \citet{aitsahalia01}, \citet{beskosetal06} 
and \citet{mijatovicschneider07} ensure that reliable parameter inference can be made 
without explicit formulae for transition densities.
An empirical application based on the joint time series
of the S\&P 100 and the  VXO implied volatility indices
reveals that our 
framework offers statistically significant advantages out of sample over extant model 
specifications in predicting implied as well as realized variance over several forecasting horizons.
Furthermore we find that the size and sign of the variance
risk premia implied by our model coincide with the model-independent results 
in~\citet{carrwu07}.

The paper is organized as follows. In Section~\ref{sec:framework}
we describe the main theoretical construction (see Theorem~\ref{thm:main}) 
for the framework we consider. Section~\ref{sec:application}
describes the nonlinear stochastic volatility model, its likelihood function and
the estimation algorithm which is used to find the parameter values implied by the
time series of the
S\&P 100 and the VXO index data. The empirical results are discussed in Subsection~\ref{subsec:empiricalresults}
and can be found in the Appendix.
Section~\ref{sec:conclusion}
concludes the paper. 


\section{The modelling framework}
\label{sec:framework}

In this section we describe the theoretical basis for the modelling framework
used in the present paper.
As mentioned in the introduction, Theorem~\ref{thm:main}
allows us to define our model under the pricing measure 
$\Qmeas$
and perform Girsanov's measure change to obtain 
any desired model 
under the physical 
measure~$\Pmeas$
in a wide class of
It\^o processes
where the state vector satisfies 
a possibly nonlinear SDE.
Theorem~\ref{thm:main}
also provides a weak solution of this SDE. 
The central building block of our approach is provided by 
the following very simple observation.

\begin{thm}
\label{thm:main}
Fix a time horizon
$T>0$
and
suppose 
$X=(X_t)_{t\in[0,T]}$
is an It\^{o} process 
with state space
$\D\subseteq \reals^{n}$ 
that
satisfies 
the following SDE
under the pricing measure 
$\QQ$
\begin{equation}
\label{eq:SDEQ}
 dX_t=\mu^{\Qmeas}(X_t)\, dt + \Sigma(X_t) \, dW_t^{\Qmeas},
 \qquad
 X_0=x_0\in\D,
\end{equation}
where the drift is given by the function
$\mu ^{\Qmeas}:\D\rightarrow\reals^{n}$
and
$W^{\Qmeas}=(W^{\Qmeas}_t)_{t\in[0,T]}$
is a standard 
$n$-dimensional Brownian motion
under 
$\QQ$.
We further assume that the volatility function 
$\Sigma: \D\rightarrow \reals^{n\times n}$
satisfies 
$|\det \Sigma(x)|>0$
for all
$x\in\D$.
Let
$f:\D\to\RR^n$
be any measurable function
with coordinates
$f_j:\D\to\RR$,
$j=1,\ldots,n$,
and
define the function 
$D:\D\rightarrow \reals_{+}$
by the formula
$$D(x):= 
\exp\left[-\frac{c}{|\det \Sigma(x)|}-c\sum_{j=1}^n |f_j(x)|\right]
$$ 
where
$c$
is some positive constant. 
Then the function
$\Lambda:\D\to\RR_+$,
defined by the formula
$$\Lambda(x):=D(x)\Sigma^{-1}(x)f(x),$$
is  bounded and the process 
$\eta=(\eta_t)_{t\in[0,T]}$
given by
$$\eta_t =\exp \brac{\int_{0}^{t}\Lambda(X_s)dW^\QQ_s- 
\frac{1}{2}\int_{0}^{t}\Lambda(X_s)^{\T}\Lambda(X_s)ds}, \quad t \leq T,$$
is a  
$\QQ$-martingale.
Then the dynamics of
$X=(X_t)_{t\in[0,T]}$ 
under 
the real world measure
$\PP$,
which is defined via the Radon-Nikodym derivative
$\frac{d\PP}{d\QQ}=\eta_T$,
are given by
\begin{equation}
\label{eq:Pproc}
 dX_t=(D(X_t)f(X_t) + \mu ^{\Qmeas}(X_t))\, dt + \Sigma(X_t) \, dW_t^{\Pmeas},\qquad X_0=x_0,
\end{equation}
where 
$W^{\Pmeas}=(W^\PP_t)_{t\in[0,T]}$
is a standard $n$-dimensional Brownian motion under the measure
$\Pmeas$,
defined by 
$W^\PP_t:=W^\QQ_t-\int_0^t\Lambda(X_s)ds$.
\end{thm}

The proof of Theorem~\ref{thm:main} 
follows 
by construction since 
the random variable 
$\Lambda(X_t)$
is bounded uniformly in 
$t\in[0,T]$.
Therefore the Novikov criterion 
(see Proposition~1.15 in Chapter~VIII of~\cite{revuzYor99})
applies and the density process 
$\eta$
is a true martingale under the pricing measure 
$\QQ$.
The other statements in Theorem~\ref{thm:main} 
follow from Girsanov's theorem
(see Theorems~1.4 and~1.7 in Chapter~VIII of~\cite{revuzYor99}).

The sole purpose of the dampening function 
$D$ 
in Theorem~\ref{thm:main}
is to ensure the existence of the real world probability measure 
$\Pmeas$,
which is equivalent to the pricing measure 
$\QQ$.
Note that 
the positive constant 
$c$ 
in the function 
$D$
can be made arbitrarily small. 
In the case the volatility function 
$\Sigma$
and the drift function
$f$
are continuous in the state variable
$x\in\D$,
the dampening factor
$D$ 
equals one 
in a finite precision environment (i.e. a computer)
on an arbitrarily large compact subset of the domain
$\D$.
As a consequence 
we have a large amount of
freedom when specifying the drift function
$f + \mu^{\Qmeas} \approx \mu ^{\Pmeas}$
that can achieve the desired  
drift behaviour of the model 
under the real world measure
$\Pmeas$.
The key observation here is that the constant
$c$
in the function 
$D$
does not need to be estimated. 
It is enough to know that it exists.
This 
by Theorem~\ref{thm:main} implies 
that
the solution of SDE~\eqref{eq:Pproc}
also exists and that
the corresponding process
behaves in the desired way 
under the real world measure.

It remains to specify a flexible model under the pricing measure 
$\Qmeas$. 
We should stress here that the only assumption 
in Theorem~\ref{thm:main}
on the process
$X$
under the pricing measure 
$\QQ$
is that it exists and satisfies SDE~\eqref{eq:SDEQ} in the theorem. 
Therefore the specification of the measure 
$\QQ$
is in practice informed by the analytical tractability of the
model in terms of the pricing of derivatives.
A common choice in 
the multivariate diffusion setting are affine processes. 
The existence of this class of models
is established in~\citet{duffiefilipovicschachermayer03} 
and the algorithms for the pricing of contingent claims, which rest
on the extended transform methods, are developed in~\citet{duffiepansingleton00}. 
In the application discussed in this paper (see Section~\ref{sec:application}) 
we shall deviate slightly from the affine class
and consider a stochastic volatility model based on a
GARCH diffusion, 
which is in the class of polynomial models. 
The existence of the process is not difficult to prove and
will be established in Section~\ref{sec:application}.
In this model it is possible to compute analytically
an approximation for the implied volatility in terms 
of the model parameters under 
$\QQ$.
This feature is crucial because the goal is to estimate
the risk-neutral and the real world parameters simultaneously.
We conclude this section with a simple affine example that illustrates 
the application of Theorem~\ref{thm:main}.

\paragraph{Example:} Consider the following univariate short rate model
 \begin{equation*}
  	dr_t=(a^{\Qmeas} - b^{\Qmeas}\, r_t)\, dt + \sigma \sqrt{r_t}dW^{\Qmeas}_t,
 \end{equation*}
 with state space
 $\D=(0,\infty)$
 and 
 $2a^{\Qmeas}>\sigma^2$.
  Let
\begin{gather*}
 f(r)=a^{\PP}- a^{\Qmeas} - (b^\PP -b^{\Qmeas})r\quad \text{and} \quad
 D(r)=e^{-c\left(r + \frac{1}{\sqrt{r}}\right)}.
\end{gather*}
Under the real world probability measure 
$\Pmeas$
the process satisfies the SDE
\begin{equation*}
 dr_t = \brat{a^{\Qmeas} - b^{\Qmeas}\, r_t + e^{-c\left(r_t + \frac{1}{\sqrt{r_t}}\right)} (a^\PP - a^{\Qmeas} - (b^\PP -b^{\Qmeas})r_t)}dt + \sigma
 \sqrt{r_t}dW^{\Pmeas}_t.
\end{equation*}
Since the domain 
$\D$
of the process
$(r_t)_{t\in[0,T]}$ 
is the positive real line,
we can choose the constant
$c$  
small enough
so that 
for numerical purposes such as estimation we can assume that the dynamics of 
the process is given by 
\begin{equation*}
 dr_t=\brac{a^\PP - b^\PP \, r_t}dt + \sigma \sqrt{r_t}dW^{\Pmeas}_t.
\end{equation*}


\section{Application}\label{sec:application}

In this section we are going to apply Theorem~\ref{thm:main} 
in order to estimate an affine (linear) and a nonlinear model on the joint time series of the 
S\&P 100 and VXO implied volatility index. We first describe the data set
and the two models that will be used for prediction and then
discuss in some detail the expected maximum 
likelihood (EML) estimation algorithm,
which is used for parameter inference of the nonlinear 
diffusions. Finally we perform a statistical test given in~\cite{clarkwest07} 
on the estimated models with respect to their forecasting power.
\subsection{Data}
\label{subsec:Data}
\begin{figure}
 \begin{small}
  \begin{center}
  \input{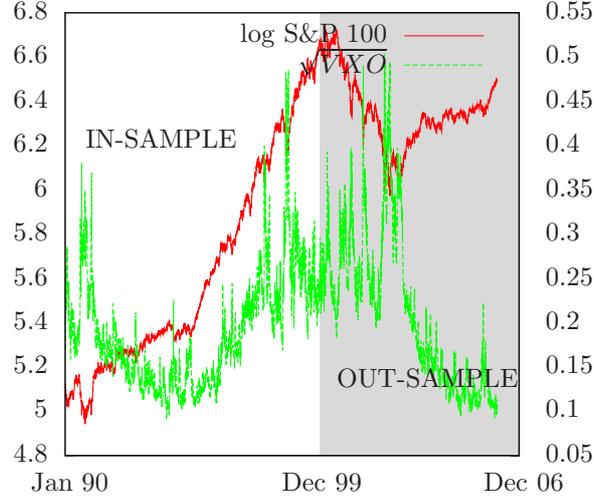}
    \caption{\label{fig:spxandvxo}{{\bf Log of S\&P 100 index and VXO: } The figure shows the evolution of 
    the logarithm of the S\&P 100 index (left y-axis) and of the implied volatility index VXO (right y-axis). 
    The sample is comprised of an in-sample period (shaded in white) and an out-sample period (shaded in grey).} }
\end{center}
  \end{small}
 \end{figure}

Models are estimated using daily S\&P 100 log prices 
and daily VXO implied volatilities. The VXO index is defined in terms of
the current value of the expected realised variance of 
S\&P 100 over a period of one month. 
The \href{http://www.cboe.com}{CBOE} computes the value of VXO
using a carfully designed portfolio of  exchange traded call and put options
on the S\&P 100 
that expire in one month's time.
The algorithm used by \href{http://www.cboe.com}{CBOE} enables them to obtain a 
time series of model independent implied volatility.
Figure \ref{fig:spxandvxo} shows the trajectory of the VXO implied volatility index published by 
the \href{http://www.cboe.com}{CBOE} and the logarithm of the S\&P 100 index. 

We partition our data set into two non-overlapping subsets. 
The \emph{in-sample} period ranges from 2 January 1990 until 31 December 1999 
and the \emph{out-of-sample} period lasts from 3 January 2000 until 29 December 2006. 

\subsection{S\&P 100 stochastic volatility model}\label{subsec:model}
We choose a stochastic GARCH diffusion variance model for the joint times series of the logarithm of the  
S\&P 100 prices and instantaneous variance,
which evolves under the  pricing measure 
$\Qmeas$
according to the SDE
\begin{eqnarray}
 	dX_t&=&(r-\frac{1}{2}V_t)\, dt + \rho \sqrt{V_t} dW^{V\Qmeas}_t + \sqrt{1 - \rho ^2}\sqrt{V_t}dW^{X\Qmeas}_t, \label{eq:sqlin}\\
 	dV_t &=& (b_0^\QQ + b_1^\QQ\, V_t)dt + \sigma V_t \,  dW^{V\Qmeas}_t\label{eq:vqlin},
\end{eqnarray}
where 
$W^{V\Qmeas}=(W^{V\Qmeas}_t)_{t\in[0,T]}$
and
$W^{X\Qmeas}=(W^{X\Qmeas}_t)_{t\in[0,T]}$
are two independent standard Brownian motions.
Note that the risk-neutral drift 
$\mu^\QQ$
from Theorem~\ref{thm:main}
can be expressed as
\begin{eqnarray*}
 	\mu^{\Qmeas}(V_t)&=
 	\begin{pmatrix}
 	 r-\frac{1}{2}V_t \\
 	 \kthq{} + \kq{}\, V_t
 	\end{pmatrix}.
\end{eqnarray*}
It is well known that the SDE 
in~\eqref{eq:vqlin}
has a solution for all values of
$b_0^\QQ$,
$b_1^\QQ$
and
$\sigma$.
Assume that
\begin{eqnarray}
\label{eq:Restriction_b0}
b_0^\QQ>0,
\end{eqnarray}
and note that the comparison theorem for the solutions of 
SDEs
(see Proposition~5.2.18 in~\cite{karatzasshreve93}),
applied to 
$V=(V_t)_{t\in [0,T]}$
and the geometric Brownian motion that solves~\eqref{eq:vqlin}
when
$b_0^\QQ=0$,
implies that 
the process 
$V$
does not leave the interval
$(0,\infty)$
in finite time.
It therefore follows that 
under condition~\eqref{eq:Restriction_b0}
we can define the process
$X=(X_t)_{t\in[0,T]}$
as a stochastic integral given by~\eqref{eq:sqlin}.
This argument shows that the process
$(X,V)^\T$
with state space
$\D=\RR\times(0,\infty)$
exists under the pricing measure 
$\QQ$
and that it follows SDE~\eqref{eq:sqlin}--\eqref{eq:vqlin}.
It is shown in~\cite{formansorensen08}
that if in addition we have
$b_1^\QQ<0$,
then the variance process 
$V$
is ergodic.

The power in the volatility function of SDE~\eqref{eq:vqlin}
(i.e. the CEV power) for the GARCH
diffusion 
$V$
is equal to one.
This is 
a pragmatic and parsimonious educated guess between the CEV powers of 0.65 in \citet{aitsahaliakimmel05b} 
and 1.33/1.17 in \citet{jones03b} 
for similar data sets.
More sophisticated volatility-of-volatility functions are also possible, e.g.
$(\beta_0+\beta_1 \, V_t + \beta _2 \, V_t^{\beta_3})^{1/2}$, 
but the existence of solutions of such SDEs 
is more difficult to establish. 
Here we use the simple GARCH diffusion process (see~\cite{nelson90}
for this terminology) because our focus in this paper is on the 
nonlinear drift specification.

Under the physical measure
$\PP$
we shall consider a linear and a nonlinear drift specifications.
The former 
$\Pmeas$-drift is linear in the state variables (LN model)
and is given by the formula
\begin{equation}
	\mu ^{\Pmeas}_{\text{LN}}(V_t):=
 \begin{pmatrix}
  a_0 + a_1 \, V_t \\
  \kthq + b_1 \, V_t
 \end{pmatrix}\label{eq:lnpdrift}.
\end{equation}
This drift corresponds to a usual market price of risk assumption as stated in~\citet{jones03b} and 
in~\citet{aitsahaliakimmel05b}. 
In the language of Section~\ref{sec:framework}
the drift 
$\mu ^{\Pmeas}$
can be expressed using the function 
$f$
given by
\begin{equation}
\label{eq:f_lin}
 f_{\text{LN}}(V_t)=
 \begin{pmatrix}
    a_0- r + (a_1+\frac{1}{2}) V_t \\
    (b_1-\kq) \, V_t
 \end{pmatrix}.
\end{equation}
The linear model LN will serve as a benchmark for  
econometric relevance of the nonlinear model (NL model) 
whose drift under the physical measure 
$\PP$
is given by
\begin{equation}\label{eq:nlpdrift}
 \mu ^{\Pmeas}_{\text{NL}}(V_t):=
 \begin{pmatrix}
  a_0 + a_1 \, V_t \\
  b_0 + b_1 \, V_t + b_2 \,  V_t^2 + b_3/V_t
 \end{pmatrix}.
\end{equation}
The corresponding function 
$f$
from Section~\ref{sec:framework}
is given as 
\begin{equation}
\label{eq:f_nonlin}
 f_{\text{NL}}(V_t)=
 \begin{pmatrix}
    a_0- r + (a_1+\frac{1}{2}) V_t \\
    b_0 - \kthq  + (b_1-\kq) \, V_t + b_2 \,  V_t^2+b_3\, / \, V_t 
 \end{pmatrix}.
\end{equation}
As described in 
Section \ref{sec:framework} 
the dampening functions 
$D_{\text{NL}}, D_{\text{LN}}$ 
can be made arbitrarily close to 
one
through the choice of the constant
$c$. 
Hence for
numerical purposes and econometric implementation it suffices to work
directly 
with the drifts 
$\mu^{\Pmeas}_{\text{NL}}$ and $\mu^{\Pmeas}_{\text{LN}}$
given in~\eqref{eq:lnpdrift}
and~\eqref{eq:nlpdrift}
respectively.

For implementation 
it is convenient to consider 
the process
$Y=(Y_t)_{t\in[0,T]}$,
given by
$Y_t:=\gamma(V_t)$,
where the transformation
$\gamma:(0,\infty)\rightarrow \reals$ 
of the variance process 
is defined by the formula
\begin{equation}\label{eq:vartransformation}
\gamma (v)=\frac{\log v}{\sigma}.
\end{equation}
The evolution of the process 
$Y$
under the physical measure 
$\PP$
is given by 
\begin{gather}
 	dY_t=\brat{\brac{\kthq{} + b_1\, V_t}\frac{1}{\sigma \, V_t}-\frac{\sigma}{2}}dt + dW^{\Pmeas}_V(t)\label{eq:tlinvar},
\end{gather}
in linear model~\eqref{eq:lnpdrift} and by 
\begin{gather}
 	dY_t=\brat{\left(b_0 + b_1\, V_k + b_2\, V_t^2+\frac{b_3}{V_k}\right)\frac{1}{\sigma \, V_k} -\frac{\sigma}{2}}dt + dW^{\Pmeas}_V(t)
	\label{eq:tnonlinvar},
\end{gather}
in nonlinear model~\eqref{eq:nlpdrift}. 

For estimation purposes we partition the parameter vector 
$\theta$
into four classes.
The first class 
$\theta ^{\sigma}$ 
contains the parameters that influence the dynamics under both the physical measure 
$\Pmeas$ 
and the pricing measure 
$\Qmeas$. 
The second class 
$\theta ^{\Qmeas}$ 
contains the parameters that arise only under the pricing
measure
$\Qmeas$. 
The third set 
$\theta ^{X\Pmeas}$ 
contains the parameters that influence the dynamics of the process 
$X$
only
under the physical measure
$\PP$
and
the fourth class 
$\theta ^{V\Pmeas}$ 
contains the parameters that arise 
only under the measure
$\PP$
in the SDE for the variance process
$V$.
It is clear that
we can express 
$\theta=\theta ^{\sigma}\cup \theta^{\Qmeas} \cup \theta ^{X\Pmeas}\cup \theta ^{V\Pmeas}$
and that these four classes are pairwise disjoint.

\begin{table}
 \begin{center}
  	\begin{tabular}{l l l l l}
  	  & $\theta ^{\sigma}$ & $\theta ^{\Qmeas}$ & $\theta ^{X\Pmeas}$ & $\theta ^{V\Pmeas}$ \\
  	  \hline \hline
  Linear Spec. \eqref{eq:lnpdrift} & $\sigma, \rho, \kthq{}$ & $\kq{}$ & $a_0, a_1$ & $b_1$ \\
  Nonlinear Spec. \eqref{eq:nlpdrift} & $\sigma, \rho$ & $\kthq{}, \kq{}$ & $a_0, a_1$ & $b_0, b_1, b_2, b_3$ \\
  	 \hline \hline
  	\end{tabular}
 \end{center}
  \caption{\label{tab:parameter}{{\bf Parameter sets for the linear and the nonlinear model: } the table displays the 
  partition of the parameter vector 
  $\theta$ 
  into the following groups: 
  $\theta ^{\sigma}$ (the parameters that 
  influence the dynamics under both the risk-neutral measure
  $\Qmeas$ 
  as well as the physical measure 
  $\Pmeas$); 
  $\theta^{\Qmeas}$
  (parameters that influences the 
  only the risk-neutral dynamics);
  $\theta^{X\Pmeas}\cup\theta ^{V\Pmeas}$
  (parameters that appear only under the physical measure~$\PP$).}}
\end{table}

\subsection{Likelihood function}
The \emph{instantaneous} stochastic variance is a latent variable
even though a time series of the \emph{implied} variance is available
through the VXO index. 
Note that the drift of the variance 
$V$
in our model,
given in SDE~\eqref{eq:vqlin}
under the pricing measure
$\QQ$,
is affine. 
Therefore the current price of the variance swap is linear
in the current value of the variance 
$V_t$
in our model as the following
simple calculation, based on Fubini's theorem, demonstrates
\begin{gather}
 	\frac{1}{\Delta}\, \evqt{t}{\int _t^{t + \Delta}V_s \, ds} = A(\theta ^{\Qmeas}, \Delta) + B(\theta^{\Qmeas}, \Delta) \, V_t, \qquad \Delta>0,
 	\intertext{where the coefficients $A(\theta ^{\Qmeas}, \Delta)$ and $B(\theta ^{\Qmeas}, \Delta)$ are given by}
 	B(\theta ^{\Qmeas}, \Delta) = 
	\frac{1}{\kq{} \Delta}\brac{\exp (\kq{} \Delta) - 1}, \quad A(\theta^{\Qmeas}, \Delta) =-\frac{\kthq{}}{\kq{}}(1-B(\theta ^{\Qmeas}, \Delta)).
\end{gather} 
We define $IV_t$  as the squared VXO index (described in Section~\ref{subsec:Data}) observed at time $t$. It is directly related to
the expected variance over the period
of 
$22$ 
days
(i.e. 
$\Delta = 22/262$)
by the formula 
\begin{equation}\label{eq:ivproxy}
 	IV_t \approx \frac{1}{\Delta}\, \evqt{t}{\int _t^{t + \Delta}V_s \, ds}.
\end{equation}
This approximation is very good and the error stems solely from the fact that the algorithm
that computes the value of VXO uses finitely many options. 

We now exploit the relationship 
in~\eqref{eq:ivproxy} to express the log-likelihood function for both the linear and the 
nonlinear model described by 
the real world drifts given in~\eqref{eq:lnpdrift}
and~\eqref{eq:nlpdrift} respectively and by 
SDE~\eqref{eq:sqlin}--\eqref{eq:vqlin}
under the pricing measure~$\QQ$.
By the Markov property property 
we can in both models
decompose the log-likelihood into a sum of 
log-transition densities 
(for ease of notation we henceforth denote
$IV_{t_i}$ 
by
$IV_i$ 
and 
$X_{t_i}$ 
by
$X_i$)
as follows
\begin{equation}\label{eq:loglikelihood}
 \ell(X_1, IV_1 , \ldots, X_N, IV_N \mid X_0, IV_0,  \theta) =  \sum _{i=1}^{N} \log p^{IV}(X_i, IV_i \mid X_{i-1}, IV_{i-1}, \theta),
\end{equation}
where
$p^{IV}(X_i, IV_i \mid X_{i-1}, IV_{i-1}, \theta)$
denotes the conditional transition density 
of the random vector
$(X_{t_i},IV_{t_i})^\T$.
The linear transformation 
\begin{eqnarray}
\label{eq:LinearTransform}
V_t = \frac{IV_t - A(\theta ^{\Qmeas}, \tau)}{B(\theta ^{\Qmeas}, \tau)},
\end{eqnarray}
which follows from~\eqref{eq:ivproxy}, 
implies that we can express the log-likelihood
as
\begin{equation}\label{eq:loglikelitrans1}
 \sum _{i=1}^{N} \log p^V(X_i, V_i \mid X_{i-1}, V_{i-1}, \theta) - N \, \log B(\theta ^{\Qmeas}, \tau),
\end{equation}
where 
$p^V(X_i, V_i \mid X_{i-1}, V_{i-1}, \theta)$
denotes the conditional transition density 
of the random vector
$(X_{t_i},V_{t_i})^\T$
given the values of 
$X_{i-1}$ 
and
$V_{i-1}$.
The final change of variable
$Y_t=\gamma(V_t)$
given in~\eqref{eq:vartransformation}  
yields the log-likelihood
which takes the form
\begin{equation}\label{eq:loglikelitrans2}
\ell(\theta)= \sum _{i=1}^{N} \brat{\log p(X_i, Y_i \mid X_{i-1}, Y_{i-1}, \theta) -\sigma Y_i } - N \, (\log B(\theta ^{\Qmeas}, \tau)-\sigma),
\end{equation}
where
$p(X_i, Y_i \mid X_{i-1}, Y_{i-1}, \theta)$
denotes the conditional transition density 
of the random vector
$(X_{t_i},Y_{t_i})^\T$.

\subsection{Limited information expected maximum likelihood estimation}\label{subsec:lieml}
The transition densities  
for the transformed variance processes~\eqref{eq:tlinvar} and~\eqref{eq:tnonlinvar} 
that arise in the log-likelihood~\eqref{eq:loglikelitrans2}
are not available in closed form. To overcome this issue and that of the supposedly flat likelihood function 
we apply expected maximum likelihood (EML) estimation algorithm from~\citet{mijatovicschneider07}. 
This technique makes use of the closeness of the law of the Brownian bridge to the true law of the diffusion
bridge, and of the Euler scheme approximation 
for the transition density when the time interval 
between observations is small. 
However, the EML algorithm cannot be directly applied to the present econometric problem, 
as it only works for the estimation of one-dimensional diffusions.
We therefore propose an efficient three-step 
limited-information maximum likelihood procedure, described below. 
The efficiency of the algorithm arises from the fact that EML can be used to express the 
globally optimal drift 
parameters 
$\theta ^{V\Pmeas \star}$ and $\theta ^{X\Pmeas \star}$ (optimal parameters are denoted with a superscript $^{\star}$) 
as complicated, yet closed-form functions of the parameters
$\theta^{\sigma}\cup \theta^{\Qmeas}$ and the data. 
In other words, for fixed values 
$\theta^{\sigma}\cup \theta^{\Qmeas}$ 
the data implies optimal parameter values
$\theta ^{V\Pmeas \star}$ 
and 
$\theta ^{X\Pmeas \star}$.  
The EML algorithm therefore effectively reduces the parameter space from 
$\theta ^{X\Pmeas} \cup \theta ^{V\Pmeas}\cup \theta^{\sigma}\cup \theta^{\Qmeas}$
to  
$\theta^{\sigma}\cup \theta^{\Qmeas}$. 
As a result
a conventional likelihood search using standard optimization techniques over equation \eqref{eq:loglikelitrans2} 
is necessary only for $\theta^{\sigma}\cup \theta^{\Qmeas}$. 

To make our processes suitable for EML estimation we first introduce $M-1$ auxiliary data $U_{i,1}, \ldots, U_{i,M-1}$ between each observed data pair 
$(X_{i}, Y_{i})^\T,(X_{i+1}, Y_{i+1})^\T$
with the convention that 
$U_{i,0}:=U_{i}:=(X_{i}, Y_i)^\T$, and $U_{i,M}:=U_{i+1}:=(X_{i+1}, Y_{i+1})^\T$. This augmentation leads to a total of $MN+1$ data pairs. To lighten notation we switch for the below equations to a single-index notation $U_k, \, k=0, \ldots , MN$. We set $\delta:=\frac{\Delta}{M}$ and write down the discretized version of the continuous-time SDE  eliminating heteroskedasticity in the innovations for the linear variance model (LN)
\begin{equation}\label{eq:lindiff}
\begin{split}
 	\frac{X_{k+1} -X_{k} - \rho \sqrt{e^{\sigma Y_k}}  \varepsilon ^{V}_{k+1}}{\sqrt{e^{\sigma Y_k}}\sqrt{1-\rho ^{2}}}&=\brat{\brac{a_0 + a_1 \, e^{\sigma Y_k}}\frac{1}{\sqrt{e^{\sigma Y_k}}\sqrt{1-\rho ^{2}}}}  \delta +  \,  \varepsilon ^{X}_{k+1} \\
 	Y _{k+1} - Y _k + \frac{\sigma}{2}\delta&= \brac{\kthq{} + b_1\, e^{\sigma Y_k}}\frac{1}{\sigma \, e^{\sigma Y_k}}\delta + \varepsilon ^{V}_{k+1},
\end{split}
\end{equation}
and the nonlinear variance model (NL)
\begin{equation}\label{eq:nonlindiff}
\begin{split}
 	\frac{X_{k+1} -X_{k} - \rho \sqrt{e^{\sigma Y_k}} \varepsilon ^{V}_{k+1}}{\sqrt{e^{\sigma Y_k}}\sqrt{1-\rho ^{2}}}&=\brat{\brac{a_0 + a_1 \, e^{\sigma Y_k}}\frac{1}{\sqrt{e^{\sigma Y_k}}\sqrt{1-\rho ^{2}}}}  \delta +  \,  \varepsilon ^{X}_{k+1} \\
 	Y _{k+1} - Y _k + \frac{\sigma}{2}\delta&= \brac{b_0 + b_1\, e^{\sigma Y_k} + b_2\, e^{2 \sigma Y_k}+\frac{b_3}{e^{\sigma Y_k}}}\frac{1}{\sigma \, e^{\sigma Y_k}} \delta + \varepsilon ^{V}_{k+1}.
\end{split}
\end{equation}
It can be seen that the difference equations \eqref{eq:lindiff} and \eqref{eq:nonlindiff} above for both, log stock prices,  as well as stochastic variance can 
be written in the form
\begin{align*}
 	g_X(U_{k+1}, U_k)&=(f^{X}_0(U_k)+f^{X}_1(U_k))\, \delta + \varepsilon^{X}_{k+1}\\
	g_{\model}(U_{k+1}, U_k)&=\sum _{l=0}^{L_\model}f^{\model}_l(U_k)\, \delta + \varepsilon^{V}_{k+1}, \, \model\in \brat{\text{LN, NL}},
\end{align*}
where the functions $g$ and $f$ are displayed in tables \ref{tab:stockemlfuncs} and \ref{tab:varemlfuncs}. For the linear variance model we have $L_{\text{LN}}=1$, and for
the nonlinear model we have $L_{\text{NL}}=3$. Innovations $\varepsilon ^{V}_k$ and $\varepsilon ^{X}_k$ are both identically independently $N(0, \delta)$-distributed
random variables for $k=1, \ldots, MN$. Note the appearance of $\varepsilon ^{V}_{k+1}$  in equ. \eqref{eq:lindiff} and \eqref{eq:nonlindiff}, respectively. This is
the reason why we need to estimate $\theta ^{V\Pmeas}$ first (the variance dynamics do not depend on the log stock price) and $\theta ^{X\Pmeas}$ subsequently,
conditional on $\theta ^{V\Pmeas\star}$.  Hence the terminology `limited information' EML estimation. For the below algorithm denote with $\theta ^{V^{\model}\Pmeas},
\model \in \brat{\text{LN, NL}}$ the parameters of the linear, respectively nonlinear variance process. If there is no ambiguity we will just write $\theta ^{V\Pmeas}$.

\paragraph{Optimizing $\theta ^{V\Pmeas}\mid \theta^{\sigma}, \theta^{\Qmeas}$: } Plugging the current values of $\theta^{\sigma}$ and $\theta^{\Qmeas}$ into eq. \eqref{eq:ivproxy} the observed data implies a time series of $Y$.  An estimate of the parameters of the transformed variance process $Y$ can now be obtained by means of EML \citep{mijatovicschneider07}. For this purpose we introduce functions $f$ and $g$ from difference equations \eqref{eq:lindiff} and \eqref{eq:nonlindiff} which are displayed in Table \ref{tab:functionsforeml}.
\begin{table}[h]
 	\begin{center}
 		\subfloat[Variance drift functions]{\label{tab:varemlfuncs}\begin{tabular}{l l l}
 		& $\model=\text{LN}$ & $\model=\text{NL}$ \\
 		 	\hline \hline
 		$f_{0}^{\model}(u_k)$ & $1/(\sigma e^{\sigma y_k})$& $1/(\sigma e^{\sigma y_k})$ \\
 		$f_{1}^{\model}(u_k)$ & $1/\sigma $& $1/\sigma$ \\
 		$f_{2}^{\model}(u_k)$ &  & $e^{\sigma y_k}/\sigma $ \\
 		$f_{3}^{\model}(u_k)$ &  & $1/(\sigma \, e^{2 \sigma y_k}) $ \\
 		\hline
 		$g_{\model}(u_k, u_{k-1})$ & $y_k - y_{k-1} + (\frac{1}{2}\sigma - \kthq{})\delta$ & $y_k - y_{k-1} + \frac{1}{2}\sigma\delta$\\
 		\hline \hline
 		\end{tabular}} \\
 		\subfloat[Stock drift functions for model $\model\in\{\text{NL, LN}\}$]{\label{tab:stockemlfuncs}\begin{tabular}{l c}
 		&  \\
 		 	\hline \hline
 		$f_0^X(u_k)$ & $1/(\sqrt{1-\rho^2}\sqrt{e^{\sigma y_k}})$  \\
 		$f_1^X(u_k)$ & $\sqrt{e^{\sigma y_k}}/\sqrt{1-\rho^2}$  \\
 		\hline
 		$g_X(u_k, u_{k-1})$ & $\brac{x_{k} -x_{k-1} - \rho \sqrt{e^{\sigma y_{k-1}}}\varepsilon ^{V}_{k}}/\brac{\sqrt{1-\rho^2}\sqrt{e^{\sigma y_{k-1}}}}$\\
 		\hline \hline
 		\end{tabular}}
 	\end{center}
 	 \caption{\label{tab:functionsforeml}{{\bf Function specification for EML estimation: } The tables contain the functions that
	 appear as summands in the respective drifts of the LN and NL models, which need to 
	 be evaluated in the conditional expectations in~\eqref{eq:mymatrix} and~\eqref{eq:myvector}, expressed in terms of the variable
	 $u_k:= (x_k , y_k)$.}
}
\end{table}
 For a given variance model $\model$ we put the variance drift parameters in a vector $x^{\model}:=\begin{pmatrix} b_{ 0}^{\model} , \ldots, b^{\model}_{ _{L_{\model}}}\end{pmatrix}^{\T}$. EML then yields the optimal drift coefficients $\theta^{V^{\model}\Pmeas \star}$ as the unique solution of the linear system $x^{\model}=(\varXi^{\model})^{-1}\, \varpi^{\model}$ with
\begin{gather}
\varXi^{\model}=\delta \, \lsum {n=1}{N-1} \lsum{m=0}{M-1} \begin{pmatrix}
     \evind{\dblaw{U_{n}}{U_{n+1}}}{f_{0}^{\model}(U_{n,m})f_{0}^{\model}(U_{n,m})} & \cdots & \evind{\dblaw{U_{n}}{U_{n+1}}}{f^{\model}_{{L_\model}}(U_{n,m})f^{\model}_{0}(U_{n,m})} \\
   \vdots & \ddots & \vdots \\
    \evind{\dblaw{U_{n}}{U_{n+1}}}{f^{\model}_{{0}}(U_{n,m})f^{\model}_{{L_\model}}(U_{n,m})} & \cdots & \evind{\dblaw{U_{n}}{U_{n+1}}}{f^{\model}_{{L_\model}}(U_{n,m})f^{\model}_{{L_\model}}(U_{n,m})}
  \end{pmatrix}, \label{eq:mymatrix} \\
  \varpi^{\model} = \lsum {n=1}{N-1}\lsum{m=0}{M-1}\begin{pmatrix}
        \evind{\dblaw{U_{n}}{U_{n+1}}}{g(U_{n,m+1}, U_{n,m})\, f^{\model}_{0}(U_{n,m})} \\
       \vdots \\
 \evind{\dblaw{U_{n}}{U_{n+1}}}{g(U_{n,m+1}, U_{n,m})\, f^{\model}_{{L _\model}}(U_{n,m})} 
      \end{pmatrix} \label{eq:myvector}.
\end{gather}
The symbol $\dblaw{x}{y}$ denotes the (unknown) law of the diffusion bridge pertaining to model $\model$ conditioned on the endpoints $x$ and $y$, respectively. We approximate the law of the true diffusion bridge $\dblaw{x}{y}$ with the law of a Brownian bridge $\dbrlaw{x}{y}$. It is shown in \citet{mijatovicschneider07} that $\dblaw{x}{y}$ is absolutely continuous with respect to $\dbrlaw{x}{y}$, and that there is in fact very little deviation between the two even for long time intervals. Exact draws from the Brownian bridge are obtained from the stochastic difference equation \citep{strameryan07}
\begin{equation}\label{eq:modbbridge}
 U_{i-1, m+1}=U_{i-1, m}+\frac{U_{i-1, M} - U_m}{M-m}+
  \sqrt{\frac{M-m-1}{M-m}} \, \varepsilon_{i-1, m+1} ,
\end{equation}
with $\varepsilon_{i, m}\sim N(0,\delta),\, i=1, \ldots , N-1, m=1,\ldots,M-1 $.
\paragraph{Optimizing  $\theta ^{X\Pmeas} \mid \theta^{\sigma}, \theta^{\Qmeas},  \theta ^{V\Pmeas \star}$: } Conditional on the optimal variance drift parameters $\theta ^{V\Pmeas \star}$ the $f$ and $g$ functions (the $g$ function depends on the drift parameters of the variance through $\varepsilon ^{V}$) from Table \ref{tab:varemlfuncs} can now be swapped with the functions from Table \ref{tab:stockemlfuncs}  to estimate optimal stock drift parameters $\theta^{X\Pmeas \star}$ through the solution of the linear system \eqref{eq:mymatrix} -- \eqref{eq:myvector}.

There is no direct EML estimator for $\theta^{\sigma} \cup \theta^{\Qmeas}$. To find $\theta^{\sigma\star}$ and  $\theta^{\Qmeas \star}$ we therefore need to perform a conventional likelihood search using likelihood \eqref{eq:loglikelitrans2} as the objective function. Since for any value of $\theta^{\sigma}$ and $ \theta^{\Qmeas}$ EML yields optimal $\theta ^{X\Pmeas \star}$ and $\theta ^{V\Pmeas \star}$, we see likelihood function \eqref{eq:loglikelitrans2} only as a function of $\theta^{\sigma}, \theta^{\Qmeas}$, and the data. To approximate the unknown transition densities which appear in \eqref{eq:loglikelitrans2} we use the simulation-based estimator from \citet{pedersen95} in connection with the Brownian bridge importance sample from \citet{durhamgallant02}

\begin{equation}\label{eq:sml}
 \sum _{i=1}^{N} \log p^{\model}(X_i, Y_i \mid X_{i-1}, Y_{i-1}, \theta) \approx \sum _{i=1}^{N}\log \brat{\frac{1}{S}\sum _{s=1}^{S}\frac{\prod_{m=1}^{M}p^{E\model}(U_{i-1, m} \mid U_{i-1, m-1}, \theta )}{\prod_{m=1}^{M-1}q(U_{i-1, m}\mid U_{i-1, m-1},U_{i-1, M})}}.
\end{equation}
Here, $p^{\model}$ refers to the true transition density arising from Heston dynamics with variance specification \eqref{eq:tlinvar} (LN) and \eqref{eq:tnonlinvar} (NL), respectively. The density $p^{E\model}$ denotes a normal distribution arising from the Euler discretization of the corresponding SDE. Auxiliary state variables $U_{i-1, m}, \ldots, U_{i-1, m}, \, i=1, \ldots , N, m=1,\ldots,M-1$ are simulated according to the stochastic difference equation
\begin{gather}
 	U_{i-1, m+1}=U_{i-1, m}+\frac{U_{i-1, M} - U_m}{M-m}+
  \sqrt{\frac{M-m-1}{M-m}} \, \Sigma(U_{i-1, m})\, \varepsilon_{i-1, m+1} \label{eq:modbridge},
 	\intertext{where}
 	\Sigma (U_{i-1, m})=
 	\begin{pmatrix}
 	 	\sqrt{1-\rho ^2} e^{\sigma Y_{i-1, m}} & \rho \sigma e^{\sigma Y_{i-1, m}} \\
 	 	0 & 1\end{pmatrix}, \,
 	\varepsilon_{i-1, m+1} =
 	\begin{pmatrix}
 	 \varepsilon ^{X}_{i-1, m+1} \\
 	 \varepsilon ^{V}_{i-1, m+1}
 	\end{pmatrix} \label{eq:diffmat}.
\end{gather}
Both $p^{E\model}$ and $q$ are multivariate normal densities:
\begin{gather*}
q(U_{i-1, m+1}\mid U_{i-1, m},U_{i-1, M})=\phi\brac{U_{i-1, m+1}; U_{i-1, m}+\frac{U_{i-1, M} - U_m}{M-m}, \frac{M-m-1}{M-m} \, \Sigma(U_{i-1, m})\Sigma(U_{i-1, m})^{\T}\delta} \\
p^{E\model}(U_{i-1, m+1} \mid U_{i-1, m}, \theta ) 
= \phi\brac{
\begin{pmatrix}
 X_{i-1,m+1} - X_{i-1,m} \\
g_\model(U_{i-1, m+1}, U_{i-1, m})
\end{pmatrix};
\begin{pmatrix}
 a_0 + a_1 e^{\sigma Y_{i-1, m}} \\
 \sum _{l=0}^{L_\model}f_l^{\model}(U_{i-1, m})
\end{pmatrix}\delta, \Sigma(U_{i-1, m})\Sigma(U_{i-1, m})^{\T}\delta }.
\end{gather*}
Following \citet{strameryan07} we set $S=M^2=576$. Note that the $\varepsilon$ variates appearing in \eqref{eq:modbbridge} from EML estimation may be reused in this step.


\subsection{Empirical results}\label{subsec:empiricalresults}
We assess the quality of the linear and nonlinear models introduced in Section \ref{subsec:model} by investigating 
forecasts of realized variance, implied variance and stock returns for various maturities. 
The forecasting exercise is performed both in and out of sample. 
For the out-of-sample period the model is re-estimated 
each time a new datapoint is added.
Figure \ref{fig:spxandvxo} gives a visual 
impression of the in-sample 
as well as the out-of-sample period. The corresponding estimation paths for a selection of parameters can be seen in Figures~\ref{fig:kthqpath} to~\ref{fig:rhopath}.

Point estimates and standard errors for the parameters of the nonlinear model 
NL (cf. \eqref{eq:tnonlinvar}) and the linear model LN (cf. \eqref{eq:tlinvar}) 
obtained from the limited information EML algorithm described in Section~\ref{subsec:lieml}
can be found in Table~\ref{table:parmest}. 
These parameter estimates are based on the entire sample. 
The $\Qmeas$ mean-reversion parameter $\kq$  is large and positive for the linear and the nonlinear model.
This is consistent with the explosive coefficients estimated in \citet{jones03b} and \citet{pan02}
and with the negative variance risk premia observed in \citet{carrwu07}. 
The positive estimates result in a time series for instantaneous variance that is 
located consistently below the time series of implied variance through relation \eqref{eq:ivproxy} 
(see Figure~\ref{fig:vxo}). The correlation and diffusion parameters 
$\rho$ 
and 
$\sigma$ as well as 
$\kthq$ are also comparable in scale for both specifications. However under the physical 
measure
$\Pmeas$ 
the linear and the nonlinear specifications predict different behaviour. 
Figure~\ref{fig:driftzoom} 
shows that during calm times (in the region between 0.01 and 0.04) the nonlinear specification predicts that the
instantaneous variance behaves as a random walk or a process that diverts at an even faster rate. 
Figure~\ref{fig:drift}
suggests that
there is a strong pull away from the zero boundary and from very high values in the case of the nonlinear drift.
Such behaviour cannot be reproduced with a linear drift specification 
(cf. also the drift function estimated
from the time series of the VIX index
in~\citet{bandireno09}, 
which is of a shape similar to that of the drift function in Figure~\ref{fig:drift}). 

Recall that the \textit{risk premium}
(i.e.the \textit{market price of risk})
at time 
$t\in[0,T]$
in the model 
$\model\in\{\text{LN, NL}\}$
is given by
\begin{eqnarray}
\label{eq:MArketPriceRisk}
\Lambda_\model (V_t) & = & \Sigma(V_t)^{-1} f_\model(V_t),\qquad\text{where}\quad
\Sigma(V_t) = \sqrt{V_t}
\begin{pmatrix}
\sqrt{1-\rho^2} & \rho\\
0 & \sigma \sqrt{V_t}
\end{pmatrix}
\end{eqnarray}
and 
$f_\model$
is defined in~\eqref{eq:f_lin}
and~\eqref{eq:f_nonlin}
for
$\model=\text{LN}$
and
$\model=\text{NL}$
respectively. 
The risk premium for the stochasticity of the variance 
is given by the second component 
$\Lambda_\model^V(V_t)$
of the market price of risk vector 
$\Lambda_\model(V_t)$.
Note that in the case
$\model=\text{LN}$,
the variance risk premium
$\Lambda_\text{LN}^V(V_t)$
is a non-zero constant given by
$(b_1-b^\QQ_1)/\sigma$.
The resulting time series of the 
risk premia 
reflect the difference in the estimated real world drifts
described in the previous paragraph: 
while the unconditional mean of the risk premium on the 
$W^{V\Qmeas}$ 
Brownian motion (see SDE~\eqref{eq:vqlin}) 
is similar for both specifications, 
the nonlinear model exhibits  
time-variability in the market prices of risk
(see Figure~\ref{fig:vxopremia}), in contrast to the constant 
risk premium, given by
$(b_1-b^\QQ_1)/\sigma$,
in the linear model.

\subsubsection{Forecasts}
In this subsection we consider, in addition to 
the nonlinear model (NL) and the linear model (LN),
a random walk martingale model (RW), where the prediction for any future 
value is taken to be the current value.
The forecasting power of the models is tested with 
2395 in-sample, and 1630 out-of-sample observations of implied variance forecasts. 
Each observation of forecast errors is comprised of a cross section of residuals pertaining 
to 1 day, 1 week, 4 weeks, 12 weeks (quarter trading year) and 26 weeks (half trading year) 
forecasting errors for stock returns and implied variance. Realized variance 
forecast errors are computed for horizons of 1 week, 4 weeks, 12 weeks and 26 weeks, 
where realized variance at time 
$t_i$ 
computed over $N$ days is defined as
\begin{equation}\label{eq:realizedvardef}
 RV_i(N):=\frac{262}{N}\sum _{j=i-N}^{i}(X_{j}-X_{j-1})^2.
\end{equation}

For the  model
$\model\in\{\text{RW, LN, NL}\}$
we compute the realized variance using the model-implied instantaneous variance, which is annualized by construction
\begin{equation}\label{eq:modelrealizedvardef}
 RV^\model_i(N):=\frac{1}{N}\sum _{j=i-N}^{i}V_{j}.
\end{equation}
Conditional expectations for the LN and NL models are computed by Monte Carlo integration using 
$2\cdot10^4$ 
paths with hourly discretization of the SDE.\footnote{With approximation~\eqref{eq:ivproxy} 
forecasts for the implied variance can be computed as linear functions of conditional instantaneous 
variance expectations. Expectations for both the linear and nonlinear model are 
evaluated by Monte Carlo integration. This is despite the availability of an analytic expression for the 
conditional expectation in the linear model so that both specifications are subject to the 
same simulation error (the same set of random numbers is used for the integration).} 
Tables \ref{table:realizedvol}, \ref{table:impliedvol} 
and \ref{table:stock} report the mean absolute error (MAE) and the 
root mean squared error (RMSE) of the sampling distribution of 
forecasting residuals for the realized variance, the implied variance and the stock returns, respectively. 
In addition directional forecasts as well as p-values of the~\citet{clarkwest07} test statistics for nested models
are reported.

Figure~\ref{fig:vxo} indicates structural breaks in the implied variance time series. 
The in-sample period spans these regimes, and the out-of-sample period contains both very rough 
and very calm periods. Both the linear and the nonlinear specification fail to 
capture the time-dependent long-run mean of the implied variance. The \citet{clarkwest07}
statistics indicate that both the linear, and the nonlinear model have significant advantages over the random walk. These results are in line with the risk premia approach in \citet{chernov07}. The statistics furthermore indicate that the nonlinear model has a statistically (highly) significant advantage 
in  forecasting over the linear model. This observation holds for the realized 
and implied variance for all forecasting horizons, both in sample and out of sample. 
A tremendous improvement 
for the realized variance over the RW specification can be attributed to mean reversion, which the NL and the LN model 
accommodate. The forecast residuals for all three competing models are heavily negatively skewed, however, 
and the distribution is fat-tailed as a consequence of a few heavy outliers. Table~\ref{table:realizedvol} 
additionally reports normalized MSE (NMSE) for comparison with the results in~\citet{sizova09}. 
For stock returns excellent in-sample results, most likely obtained through explicit modeling of 
the leverage effect, cannot be reproduced out of sample. The likely reason is a change in the drift regime 
(cf. Figure \ref{fig:spxandvxo}), which is not accounted for by the variance modeling.

The results for the directional forecasts of the realized and implied variance and of stock returns
are mixed. The NL and LN models show very good results 
in sample and out of sample for the realized variance. For the implied variance the directional 
forecasts are slightly worse than 
the ones in~\citet{ahoniemi06}, where forecasts are given for one maturity only.  
Directional out-of-sample 
forecasts for stock returns also suffer from the change in the drift during the out-of-sample period
mentioned above (cf. Figure \ref{fig:spxandvxo}).

\section{Conclusion}\label{sec:conclusion}
We introduce a simple continuous-time diffusion framework that combines semi-analytic pricing formulae 
with flexible nonlinear time series modeling. Using an econometrically inconspicuous dampening function 
we ensure that a solution to the nonlinear stochastic differential equation
under the physical measure exists.
We estimate a nonlinear stochastic volatility model on the joint time series of the S\&P 100 and 
the VXO implied volatility index. Forecast tests show that the nonlinear model has superior 
forecasting power over the random walk and the linear model
for short prediction horizons; 
the results  are statistically significantly better both in sample and out of sample. 
This suggests that a 
nonlinear specification of the drift under the physical measure
could potentially be very useful in trading and risk management. 

%

\bibliographystyle{apalike}
\bibliography{../masterbib/trunk/master}


\begin{appendix}


 	\newpage
 	\section{Figures and Tables}\label{app:figandtab}
 	\begin{table}[ht]
 	\begin{small}
 	\begin{center}
 		\begin{tabular}{c c c}
 		&Linear & Nonlinear \\
 		\hline \hline
 		$\sigma$ & 2.2047 & 2.1734\\
 		 & (0.0374) & (0.0554) \\
 		 & & \\
 		$\rho$ &-0.6768 & -0.6803 \\
 		 & (0.01577) & (0.0165)\\
 		 & & \\
 		$\kthq{}$ & 0.05817& 0.0500\\
 		 & (0.0046) & (0.0098)\\
 		 & & \\
 		$\kq{}$ &10.9858 & 11.3260\\
 		 & (0.00648) & (0.08008) \\
 		 & & \\
 		$a_0$ &0.0748 & 0.0284\\
 		 & (0.0308) & (0.0359)\\
 		 & & \\
 		$a_1$ & 3.3370& 6.0870 \\
 		 & (0.0119) & (0.0622) \\
 		 & & \\
 		$b_0$ &  & -0.1064 \\
 		 &  & (0.0221)\\
 		 & & \\
 		$b_1$ & -1.7645 & 8.9591 \\
 		 & (0.0355) & (0.0689) \\
 		 & & \\
 		$b_2$ & & -180.7473\\
 		 &  & (0.22490) \\
 		 & & \\
 		$b_3$ & &  0.00068\\
 		 &  & (0.00016)\\
 		 \hline \hline
 		\end{tabular}
 	\end{center}
 	\end {small}
 	\caption{\label{table:parmest}{{\bf Parameter Estimates:} The table displays parameter estimates for the linear model~\eqref{eq:tlinvar} 
	and the nonlinear model~\eqref{eq:tnonlinvar}. Huber (sandwich) standard errors are computed from the asymptotic covariance matrix 
	pertaining to the likelihood in~\eqref{eq:loglikelitrans2} with the transition density approximation in in~\eqref{eq:sml}. 
	The asymptotic covariance matrix of the estimated parameter vector $\hat{\theta}$ is computed according to the formula in
        \citet[][page 145, formula 5.8.7]{hamilton94}.
}}
\end{table}

\begin{figure}[ht]
 \begin{tiny}
  \begin{center}
  \subfloat[VXO]{\label{fig:vxo}\input{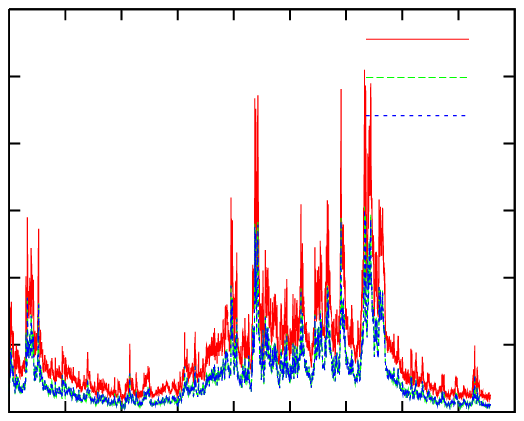}}
  \subfloat[Estimated Drift]{\label{fig:drift}\input{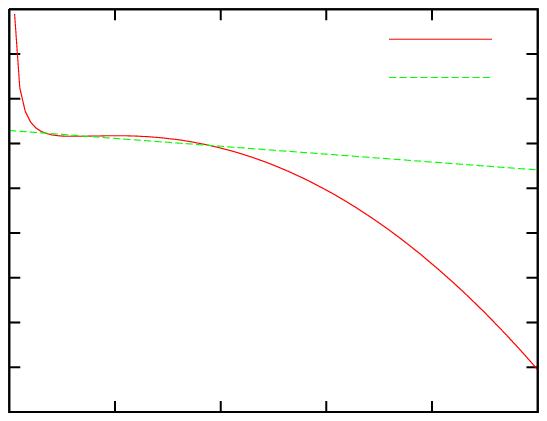}} \\
  \subfloat[Variance Premia]{\label{fig:vxopremia}\input{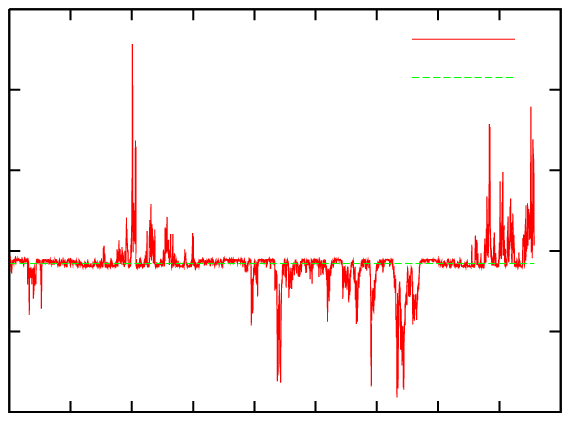}}
  \subfloat[Estimated Drift (Zoom)]{\label{fig:driftzoom}\input{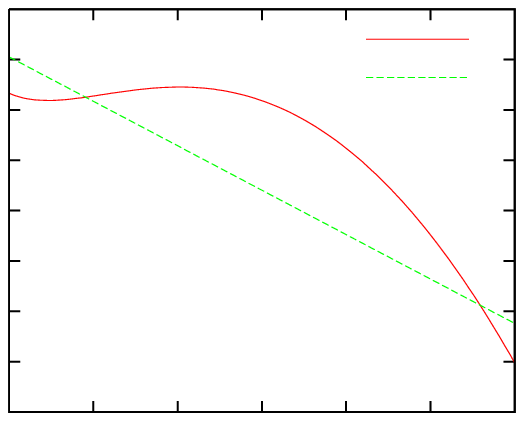}}
  \end{center}
    \caption{\label{fig:vxoandrift}{{\bf VXO and instantaneous variance: } Figure \ref{fig:vxo} displays the VXO along with the 
    instantaneous variance implied by the $\Qmeas$ parameters from Table~\ref{table:parmest}. 
    Figure~\ref{fig:drift} displays the nonlinear drift at the parameter estimates for variance model~\eqref{eq:nlpdrift}, 
    and the linear drift for model \eqref{eq:lnpdrift}. The implied time series for the risk premia of the stochastic variance,
    given by the second coordinate of the market price of risk vector~\eqref{eq:MArketPriceRisk},
    in the linear and the nonlinear model is displayed in Figure~\ref{fig:vxopremia}.}
}
  \end{tiny}
 \end{figure}

 \newpage

\begin{figure}[ht]
 \begin{tiny}
  \begin{center}
  \subfloat[$\widehat{\kthq} _n$]{\label{fig:kthqpath}\input{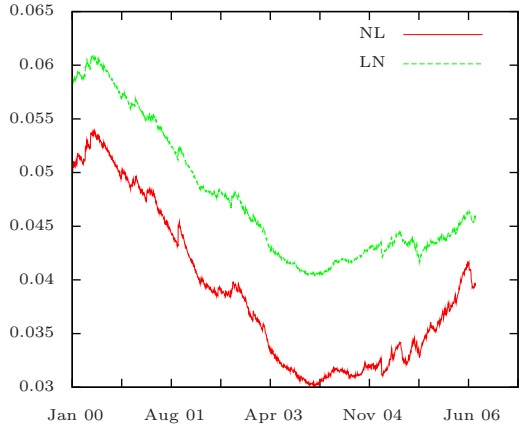}}
  \subfloat[$\widehat{\kq} _n$]{\label{fig:kqpath}\input{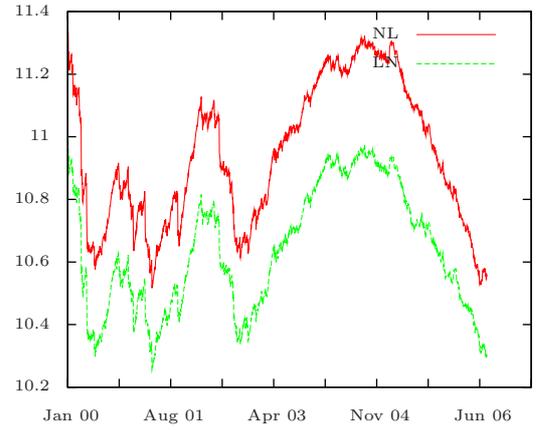}} \\
  \subfloat[$\widehat{\sigma} _n$]{\label{fig:sigpath}\input{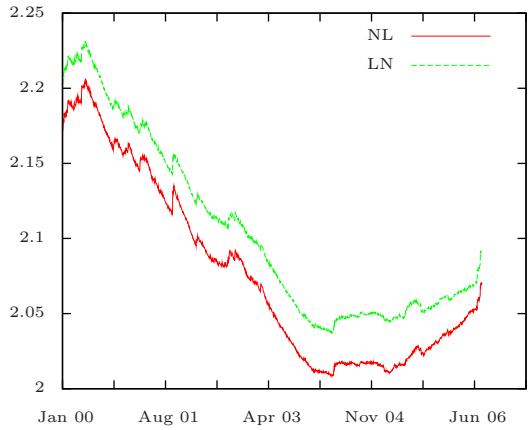}}
  \subfloat[$\widehat{\rho}_n$]{\label{fig:rhopath}\input{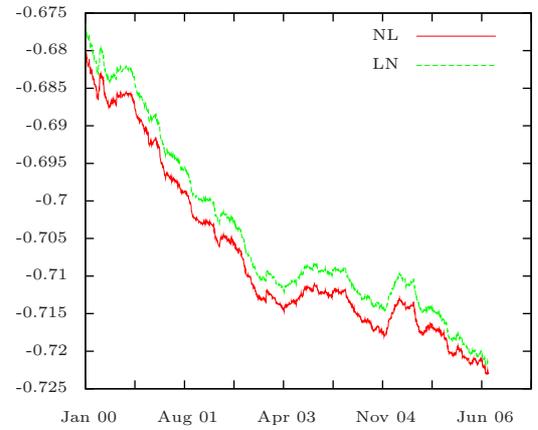}}
  \end{center}
    \caption{\label{fig:parmpaths}{{\bf Parameter paths: } Figure \ref{fig:parmpaths} displays  
    estimates for the parameters $\kthq, \kq, \sigma$, and $\rho$, respectively, as the sample 
    window is updated on a daily basis. For each date on the x-axis model \eqref{eq:tlinvar} 
    and \eqref{eq:tnonlinvar} are re-estimated using the EML methodology from Section \ref{subsec:lieml}.}
}
  \end{tiny}
 \end{figure}

 	\begin{table}
\begin{small}
 	\begin{center}
\begin{tabular}{ l l l l l l }
IN-SAMPLE    &        &    1w    &    4w    &    12w    &    26w    \\
\hline \hline                                           RMSE    &    RW    &    0.05201    &    0.05731    &    0.05908    &    0.05986    \\
   &    LN    &    0.02758    &    0.01825    &    0.01494    &    0.01348    \\
   &    NL    &    0.02776    &    0.01814    &    0.01433    &    0.01364    \\
   &        &        &        &        &        \\
NMSE    &    RW    &    238$\%$    &    567$\%$     &    943$\%$     &    1182$\%$    \\
   &    LN    &    66$\%$     &    57$\%$     &    60$\%$    &   59$\%$    \\
   &    NL    &    67$\%$     &    56$\%$     &    55$\%$     &    61$\%$     \\
   &        &        &        &        &        \\
MAE    &    RW    &    0.02191    &    0.02425    &    0.02480    &    0.02537    \\
   &    LN    &    0.01174    &    0.00914    &    0.00862    &    0.00991    \\
   &    NL    &    0.01172    &    0.00897    &    0.00807    &    0.00968    \\
   &        &        &        &        &        \\
DIR    &    RW    &    0.49227    &    0.51609    &    0.50230    &    0.49979    \\
   &    LN    &    0.70497    &    0.74133    &    0.68784    &    0.62599    \\
   &    NL    &    0.70121    &    0.74384    &    0.69118    &    0.61429    \\
   &        &        &        &        &        \\
CW     &  LN vs. RW      &    $0.00281^{(***)}$    &    $0.00612^{(***)}$    &    $0.00989^{(***)}$    &    $0.01087^{(**)}$    \\
     &    NL vs. RW    &    $0.00289^{(***)}$    &    $0.00638^{(***)}$    &    $0.01053^{(**)}$    &    $0.01166^{(**)}$    \\
     &   NL vs. LN     &    $1$    &    $0.26299$    &    $0.09399^{(*)}$    &    $0.09795^{(*)}$    \\
   &        &        &        &        &        \\
  OUT-SAMPLE &        &        &        &        &        \\
                                         \hline \hline                                           RMSE    &    RW    &    0.06789    &    0.07478    &    0.07767    &    0.07958    \\
   &    LN    &    0.03596    &    0.02714    &    0.02600    &    0.02585    \\
   &    NL    &    0.03690    &    0.02899    &    0.02723    &    0.02597    \\
   &        &        &        &        &        \\
NMSE    &    RW    &    192$\%$     &    361$\%$     &    546$\%$     &    751$\%$     \\
   &    LN    &    53$\%$     &    47$\%$     &    61$\%$     &    79$\%$     \\
   &    NL    &    56$\%$     &    54$\%$     &    67$\%$     &    79$\%$     \\
   &        &        &        &        &        \\
MAE    &    RW    &    0.03389    &    0.03718    &    0.03892    &    0.03946    \\
   &    LN    &    0.01857    &    0.01561    &    0.01705    &    0.01876    \\
   &    NL    &    0.01875    &    0.01606    &    0.01678    &    0.01825    \\
   &        &        &        &        &        \\
DIR    &    RW    &    0.49049    &    0.48557    &    0.49847    &    0.41989    \\
   &    LN    &    0.70534    &    0.69061    &    0.62247    &    0.57274    \\
   &    NL    &    0.70166    &    0.68447    &    0.61387    &    0.56967    \\
   &        &        &        &        &        \\
CW     &   LN vs. RW     &    $0^{(***)}$    &    $0^{(***)}$    &    $0.00014^{(***)}$    &    $0.00107^{(***)}$    \\
     &    NL vs. RW    &    $0^{(***)}$    &    $0^{(***)}$    &    $0.00017^{(***)}$    &    $0.00126^{(***)}$    \\
     &  NL vs. LN      &    $1$    &    $1$    &    $0.22124$    &    $0.09065^{(*)}$    \\ 
\hline \hline 
\end{tabular}
\caption{\label{table:realizedvol}{{\bf Realized variance forecasting:} This table displays mean absolute error MAE, given by
$\frac{1}{N-\tau}\sum_{i=\tau}^{N}\abs{\epsilon_i(\tau)}$,  
root mean squared forecast error RMSE, given by
$\sqrt{\frac{1}{N-\tau}\sum_{i=\tau}^{N}\epsilon_i(\tau)^2}$,
and normalized MSE (NMSE), defined as
$\sum_{i=\tau}^{N}\epsilon_i(\tau)^2/\left(\sum_{i=\tau}^{N}(RV_{i}(\tau) - \overline{RV}_i(\tau))^2\right)$, 
where $\epsilon_i(\tau):=RV_{i}(\tau) - \evt{t_i-\tau}{RV^\model_{i}(\tau)}$
and
$\tau\in\{5,22,66,131\}$.
The realized varieance 
$RV_{i}(\tau)$
is defined in~\eqref{eq:realizedvardef}
and the random variable
$RV^\model_{i}(\tau)$
is given in~\eqref{eq:modelrealizedvardef}
for any
$\model\in\{\text{RW, LN, NL}\}$,
where
RW denotes the random walk model and  LN (resp. NL) stands for the linear
(resp. nonlinear) model 
given in~\eqref{eq:tlinvar} (resp. in~\eqref{eq:tnonlinvar}). 
DIR shows the percentage of correct directional forecasts. 
CW denotes p-values for the \citet{clarkwest07} test for nested models. 
Asterisks $^{(***), (**), (*)}$ 
denote significance at the 1\%, 5\% and 10\% confidence level respectively.} }
\end{center}
\end{small}
\end{table}

\newpage

 	\begin{table}
\begin{small}
 	\begin{center}
\begin{tabular}{ l l l l l l l}
 IN-SAMPLE    &        &    1d    &    1w    &    4w    &    12w    &    26w    \\
\hline \hline                                                   MSE    &    RW    &    0.00762    &    0.01380    &    0.02029    &    0.02765    &    0.02967    \\
   &    LN    &    0.00760    &    0.01362    &    0.01945    &    0.02518    &    0.02745    \\
   &    NL    &    0.00748    &    0.01308    &    0.01833    &    0.02353    &    0.02688    \\
   &        &        &        &        &        &        \\
MAE    &    RW    &    0.00388    &    0.00748    &    0.01110    &    0.01456    &    0.01691    \\
   &    LN    &    0.00388    &    0.00751    &    0.01149    &    0.01656    &    0.02140    \\
   &    NL    &    0.00385    &    0.00732    &    0.01085    &    0.01517    &    0.01934    \\
   &        &        &        &        &        &        \\
DIR    &    RW    &    0.49269    &    0.49687    &    0.49310    &    0.47305    &    0.49812    \\
   &    LN    &    0.52069    &    0.53364    &    0.55328    &    0.58379    &    0.61220    \\
   &    NL    &    0.51692    &    0.53197    &    0.55161    &    0.55871    &    0.54952    \\
   &        &        &        &        &        &        \\
CW    &    LN vs. RW     &    $0.00413^{(***)}$    &    $0.00954^{(***)}$    &    $0.02119^{(**)}$    &    $0.01957^{(**)}$    &    $0.01366^{(**)}$    \\
     &    NL vs. RW    &    $0.00809^{(***)}$    &    $0.02778^{(**)}$    &    $0.06111^{(*)}$    &    $0.03683^{(**)}$    &    $0.02815^{(**)}$    \\
     &    NL vs. LN    &    $0.01307^{(**)}$    &    $0.04338^{(**)}$    &    $0.11345$    &    $0.05275^{(*)}$    &    $0.01554^{(**)}$    \\
   &        &        &        &        &        &        \\
 OUT-SAMPLE  &        &        &        &        &        &        \\
\hline \hline                                                   MSE    &    RW    &    0.00903    &    0.01789    &    0.02810    &    0.03919    &    0.04401    \\
   &    LN    &    0.00902    &    0.01781    &    0.02786    &    0.03926    &    0.04831    \\
   &    NL    &    0.00891    &    0.01699    &    0.02679    &    0.03684    &    0.04260    \\
   &        &        &        &        &        &        \\
MAE    &    RW    &    0.00513    &    0.01014    &    0.01676    &    0.02470    &    0.02678    \\
   &    LN    &    0.00514    &    0.01019    &    0.01749    &    0.02847    &    0.03826    \\
   &    NL    &    0.00507    &    0.00970    &    0.01605    &    0.02507    &    0.03148    \\
   &        &        &        &        &        &        \\
DIR    &    RW    &    0.48128    &    0.47759    &    0.45181    &    0.41866    &    0.36710    \\
   &    LN    &    0.51013    &    0.51320    &    0.51688    &    0.50460    &    0.43892    \\
   &    NL    &    0.50890    &    0.52977    &    0.53898    &    0.48803    &    0.46593    \\
   &        &        &        &        &        &        \\
CW     &   LN vs. RW     &    $0.02588^{(**)}$    &    $0.00106^{(***)}$    &    $0.01306^{(**)}$    &    $0.10687$    &    $0.41651$    \\
    &   NL vs. RW      &    $0.01775^{(**)}$    &    $0.06154^{(*)}$    &    $0.01735^{(**)}$    &    $0.01821^{(**)}$    &    $0.0328^{(**)}$    \\
     &   NL vs. LN     &    $0.02178^{(**)}$    &    $0.07032^{(*)}$    &    $0.02653^{(**)}$    &    $0.03312^{(**)}$    &    $0.04385^{(**)}$    \\ 
\hline \hline
\end{tabular}
\caption{\label{table:impliedvol}{{\bf Implied variance forecasting:} This table displays mean absolute error MAE,
given by $\frac{1}{N-\tau}\sum_{i=1}^{N-\tau}\abs{\epsilon_i(\tau)}$,
and 
root mean squared forecast error RMSE,
defined by 
$\sqrt{\frac{1}{N-\tau}\sum_{i=1}^{N-\tau}\epsilon_i(\tau)^2}$, 
where 
$\epsilon_i(\tau):=IV_{t_i + \tau} - \evt{t_i}{IV^\model_{t_i + \tau}}$
and
$\tau\in\{1,5,22,66,131\}$.
The random variable 
$IV^\model_t$
is defined as a linear transformation, 
given in~\eqref{eq:LinearTransform},
of the instantaneous 
variance in the model
$\model\in\{\text{NL, LN}\}$
and
$IV_t$
denotes the square of the VXO index at time 
$t$.
As in the previous table 
RW denotes the random walk model and  LN (resp. NL) stands for the linear
(resp. nonlinear) model 
given in~\eqref{eq:tlinvar} (resp. in~\eqref{eq:tnonlinvar}). 
DIR shows the percentage of correct directional forecasts. 
CW denotes p-values for the \citet{clarkwest07} test for nested models. 
Asterisks $^{(***), (**), (*)}$ 
denote significance at the 1\%, 5\% and 10\% confidence level respectively.} }
\end{center}
\end{small}
\end{table}

\newpage

 	\begin{table}
\begin{small}
 	\begin{center}
\begin{tabular}{ l l l l l l l}
IN-SAMPLE    &        &    1d    &    1w    &    4w    &    12w    &    26w    \\
\hline \hline                                                   MSE    &    RW    &    0.00936    &    0.02041    &    0.03966    &    0.07192    &    0.11658    \\
   &    LN    &    0.00932    &    0.02006    &    0.03702    &    0.05697    &    0.06950    \\
   &    NL    &    0.00932    &    0.02002    &    0.03693    &    0.05690    &    0.07029    \\
   &        &        &        &        &        &        \\
MAE    &    RW    &    0.00667    &    0.01540    &    0.03058    &    0.05492    &    0.09428    \\
   &    LN    &    0.00664    &    0.01501    &    0.02812    &    0.04245    &    0.05708    \\
   &    NL    &    0.00664    &    0.01495    &    0.02798    &    0.04225    &    0.05717    \\
   &        &        &        &        &        &        \\
DIR    &    RW    &    0.53698    &    0.58629    &    0.65650    &    0.79440    &    0.88048    \\
   &    LN    &    0.53698    &    0.58629    &    0.65650    &    0.79440    &    0.88048    \\
   &    NL    &    0.53698    &    0.58629    &    0.65650    &    0.79440    &    0.88048    \\
   &        &        &        &        &        &        \\
CW LN vs. RW    &        &    $0.00011^{(***)}$    &    $0.00594^{(***)}$    &    $0.00265^{(***)}$    &    $0.00012^{(***)}$    &    $0^{(***)}$    \\
CW NL vs. RW    &        &    $0.00022^{(***)}$    &    $0.00722^{(***)}$    &    $0.00417^{(***)}$    &    $0.00032^{(***)}$    &    $0^{(***)}$    \\
CW NL vs. LN    &        &    $0.06384^{(*)}$    &    $0.17865$    &    $0.03174^{(**)}$    &    $0.33946$    &    $1$    \\
   &        &        &        &        &        &        \\
OUT-SAMPLE   &        &        &        &        &        &        \\
\hline \hline                                                   MSE    &    RW    &    0.01218    &    0.02555    &    0.04865    &    0.07302    &    0.10893    \\
   &    LN    &    0.01221    &    0.02580    &    0.05075    &    0.08293    &    0.13686    \\
   &    NL    &    0.01221    &    0.02576    &    0.05062    &    0.08386    &    0.13894    \\
   &        &        &        &        &        &        \\
MAE    &    RW    &    0.00881    &    0.01852    &    0.03566    &    0.05529    &    0.08311    \\
   &    LN    &    0.00882    &    0.01866    &    0.03702    &    0.06202    &    0.10240    \\
   &    NL    &    0.00882    &    0.01865    &    0.03673    &    0.06161    &    0.10046    \\
   &        &        &        &        &        &        \\
DIR    &    RW    &    0.50460    &    0.48987    &    0.50583    &    0.49233    &    0.52732    \\
   &    LN    &    0.50153    &    0.49662    &    0.48312    &    0.46163    &    0.45734    \\
   &    NL    &    0.50460    &    0.48987    &    0.50583    &    0.49233    &    0.52732    \\
   &        &        &        &        &        &        \\
CW     &   LN vs. RW     &    $1$    &    $1$    &    $1$    &    $1$    &    $1$    \\
     &   NL vs. RW     &    $1$    &    $1$    &    $1$    &    $1$    &    $1$    \\
     &   NL vs. LN    &    $0.22935$    &    $0.3444$    &    $0.33081$    &    $0.38529$    &    $0.37186$    \\ 
\hline \hline
\end{tabular}
\caption{\label{table:stock}{{\bf Log stock forecasting:} 
This table displays mean absolute error MAE,
given by $\frac{1}{N-\tau}\sum_{i=1}^{N-\tau}\abs{\epsilon_i(\tau)}$,
and 
root mean squared forecast error RMSE,
defined by 
$\sqrt{\frac{1}{N-\tau}\sum_{i=1}^{N-\tau}\epsilon_i(\tau)^2}$, 
where 
$\epsilon_i(\tau):X_{t_i + \tau} - \evt{t_i}{X^\model_{t_i + \tau}}$
and
$\tau\in\{1,5,22,66,131\}$.
The random variable 
$X^\model_t$
represents the log stock in the model
$\model\in\{\text{NL, LN}\}$
and
$X_t$
denotes the recorded value of the logarithm
of the S\&P 100 
at time 
$t$.
As in the previous table 
RW denotes the random walk model and  LN (resp. NL) stands for the linear
(resp. nonlinear) model 
given in~\eqref{eq:tlinvar} (resp. in~\eqref{eq:tnonlinvar}). 
DIR shows the percentage of correct directional forecasts. 
CW denotes p-values for the \citet{clarkwest07} test for nested models. 
Asterisks $^{(***), (**), (*)}$ 
denote significance at the 1\%, 5\% and 10\% confidence level respectively.} }
\end{center}
\end{small}
\end{table}

\end{appendix}

\end{document}